\documentclass[11pt,reqno]{amsart}

\usepackage{amscd,amssymb,amsmath,amsthm}
\usepackage{graphicx}
\usepackage{epstopdf}
\usepackage{color}
\usepackage{cite}
\usepackage{amssymb}
\usepackage[russian]{babel}
\numberwithin{equation}{section} 

\begin{document}
\textit{╙─╩ 517.98}
\begin{center}\textbf{ TRANSLATION-INVARIANT GIBBS MEASURES FOR THE BLUM-KAPEL MODEL ON A CAYLEY TREE}
\end{center}

\begin{center}
 N.M.Hatamov\footnote{Namangan State University, 316, Uychi str., 160119, Namangan, Uzbekistan.\\
E-mail: nxatamov@mail.ru}, R.M.Khakimov\footnote{Namangan State University, 316, Uychi str., 160119, Namangan, Uzbekistan.\\
E-mail: rustam-7102@rambler.ru},
\end{center}\

In this paper we consider translation-invariant Gibbs measures for the Blum-Kapel model on a Cayley tree of order $k$. An approximate critical temperature $T_{cr}$ is found such that for $T\geq T_{cr}$ there exists a unique translation-invariant Gibbs measure and for $0<T<T_{cr}$ there are exactly three translation-invariant Gibbs measures. In addition, we studied the problem of (not) extremality for the unique Gibbs measure.\\


\section{Introduction}\

The Gibbs measure is a fundamental law determining the probability of a microscopic state of a given physical system and it plays an important role in determining the existence of a phase transition of a physical system, since each Gibbs measure is associated with one phase of the physical system, and if a Gibbs measure is nonunique, then it is said that there is a phase transition. It is well known that
the set of all limit Gibbs measures forms a nonempty convex compact subset of the set of all probability
measures and each point (i.e., Gibbs measure) of this convex set can be uniquely expanded in its
extreme points. In this connection, it is especially interesting to describe all extreme points of this convex
set, i.e., the extreme Gibbs measures (see\cite{6}-\cite{Si}).

Many papers are devoted to the study of limit Gibbs measures on a Cayley tree for such models of statistical physics as Ising model, Potts model, HC model and SOS model (see for example \cite{HN}-\cite{1}). In particular, in \cite{KRK} it was fully describe the set of translation-invariant Gibbs measures for the ferromagnetic $q$-state Potts model and it is proved that the number of translation-invariant measures can be up to
$2^q-1$ and in \cite{KR} the extremality problem is studied for these measures. In \cite {RKh1} Gibbs measures for three state HC models are studied on a Cayley tree of order $k\geq1$ and the nonuniqueness of the translation-invariant Gibbs measure is proved. Moreover, areas where the measures are (not) extreme are given. In the monograph \cite{R} the results on limit Gibbs measures can be found in more detail.

This paper is devoted to the study of the Blum Kapel model which has not yet been studied on a Cayley tree. This is two-dimensional spin system, where spin variables taking values from the set: $\Phi=\{-1,0,+1\}$. It was originally introduced to study $He^3-He^4$ phase transition (see \cite{CO}).
We can consider this model as the system of particle with spin. The value $\sigma(x)=0$ {of spin} on the lattice vertex (or on the tree node) $x$ corresponds to the absence of particles (vacancy) and values  $\sigma(x)=+1,-1$ to the presence of a particle with spin $+1,-1$ on the vertex $x$, respectively (see \cite{CO}-\cite{HK}).

This paper is organized as follows. In Sec. 2, we present the basic definitions and known facts. In
Sec. 3, we prove a theorem that ensures the condition of consistency of a measure. In Sec. 4, an approximate critical temperature $T_{cr}$ is found such that for $T\geq T_{cr}$ there exists a unique translation-invariant Gibbs measure and there are exactly three translation-invariant Gibbs measures for the considered model for $0<T<T_{cr}$. In Sec. 5 the sets where the existing single measure for $T>0$ is (not) extremality are given.

\section{Preliminary Information}\

A Cayley tree  $\Gamma^{k}=(V,L)$ of order $k\geq 1$ is an infinite tree, i.e., a graph without cycles such that each
vertex has precisely $k+1$ edges, where $V$ is the set of vertices of the graph $\Gamma^{k}$, $L$ is the set of its edges. Let $i$ be the incidence function associating each edge $l\in L$ to its endpoints $x,y\in V$. If $i(l)=\{x, y\}$, then $x$ and $y$ are called the nearest neighbors of a vertex, and we write this as ${\langle x,y\rangle}$. The distance $d(x,y), x, y \in V$ on the Cayley tree is defined as

$d(x,y)=\min\{d | \exists \ {x=x_{0}}, x_{1}, ... , x_{d-1},
x_{d}=y\in V $ such that $\langle x_{0}, x_{1}\rangle, ... ,
\langle x_{d-1}, x_{d}\rangle\}.$

We consider the model in which spin variables taking values from the set
$\Phi=\{-1,0,+1\}$. We then define a \emph{configuration} $\sigma$ on $V$
as a function $x\in V\rightarrow \sigma(x)\in \Phi$.
The set of all configurations coincides with $\Omega=\Phi^V$. Let
$A\subset V$. We denote the space of configurations defined on a set $A$ by $\Omega_A$.

The Hamiltonian of the Blum-Kapel model is given by the formula
$$H(\sigma)=-J\sum_{\langle x, y\rangle, x, y\in V; }\sigma(x)\sigma(y), $$
where $J>0$.

For a fixed $x^0\in V$ we write $x<y$ if a path from $x^0$ to $y$ runs through $x$.

We denote
$$W_n=\{x\in V : d(x^0, x)=n\}, \ V_n=\{x\in V : d(x^0, x)\leq n\}.$$

A vertex $y$ is called a \emph{"child"} of a vertex $x$ if
$x<y$ and $d(x,y)=1$.

We let $S(x)$ denote the set of \emph {"children"} of a vertex $x\in V$.

Let $h:x\mapsto h_x = (h_{-1,x},h_{0,x},h_{+1,x})$ be a vector-valued function on $x\in V\diagdown\{x^{0}\}.$ We consider the probability measure
$\mu^{(n)}$ on $\Omega_{V_n}$
\begin{equation}\label{f.1}
\mu^{(n)}(\sigma_n)=Z_n^{-1}exp\{-\beta H(\sigma_n)+\sum_{x\in W_n}h_{\sigma(x), x}\}.
\end{equation}
Here $\sigma_n\in \Omega_{V_n}$ ш $Z_n$ is a normalization factor,
$$Z_n=\sum_{\overline{\sigma}_n\in
\Omega_{V_n}}exp\{-\beta H(\overline{\sigma}_n)+\sum_{x\in
W_n}h_{\overline{\sigma}(x),x}\},$$
where $h_ {\overline{\sigma} ,x}\in R$.

The probability measure $\mu^{(n)}$ is said to be consistent if for all $n\geq 1$ and any $\sigma_{n-1}\in \Omega_{V_{n-1}}$:
\begin{equation}\label{f.2}
\sum_{\sigma^{(n)}}\mu^{(n)}(\sigma_{n-1}, \sigma^{(n)})=\mu^{(n-1)}(\sigma_{n-1}).
\end{equation}

In this case, there is a unique measure $\mu$ on $\Omega_V$
such that
$$\mu(\{\sigma\mid_{V_n}=\sigma_{n}\})=\mu^{(n)}(\sigma_n)$$
for all $n\geq 1$ and any $\sigma_n\in \Omega_{V_n}.$

\section{The system of functional equations}\

A condition for $h_{i,x}$ ensuring the consistency of the measures $\mu^{(n)}$
is formulated in the next theorem.

\textbf{Theorem 1.} \emph{Let} $k\geq 2$. \emph{Sequence of probabilistic measures} $\mu^{(n)}(\sigma_n), n=1,2, . . .$ defined by (\ref{f.1})
\emph{are consistent if and only
if the equalities}
\begin{equation}\label{f.3}\left\{
\begin{array}{ll}
z_{+1,x}=\prod_{y\in S(x)}\frac{\lambda z_{+1,y}+\frac{1}\lambda
z_{-1,y}+1}{z_{+1,y}+z_{-1,y}+1}, \\[4mm]
z_{-1,x}=\prod_{y\in S(x)}\frac{\frac{1}\lambda z_{+1,y}+\lambda
z_{-1,y}+1}{z_{+1,y}+z_{-1,y}+1},\\
\end{array}
\right.\end{equation}
\emph{where} $\lambda=exp\{J\beta\}, \beta=1/T,
z_{i,x}=exp(h_{i,x}-h_{0,x}), \ i=+1,-1$, hold for any $x\in
V$.

\textbf{Proof.} \textbf{Necessity}. By the consistency condition (\ref{f.2}) we get
\begin{equation}\label{f.4}\frac{Z_{n-1}}{Z_n}\sum_{\omega_n\in\Omega_{W_n}}\prod_{x\in
W_{n-1}}\prod_{y\in S(x)}
exp(J\beta\sigma_{n-1}(x)\omega_n(y)+h_{\omega_n(y),y})=\prod_{x\in
W_{n-1}}exp(h_{\sigma_{n-1}(x),x}),
\end{equation}
where $\sigma(x)\in\Phi$.

Fix $x\in W_{n-1}$ and consider three configurations $\sigma_{n-1}=\overline{\sigma}_{n-1}$, $\sigma_{n-1}=\widetilde{\sigma}_{n-1}$ and $\sigma_{n-1}=\widehat{\sigma}_{n-1}$ on $W_{n-1}$ which coincide on $W_{n-1}\setminus \{x\}$, and rewrite now the equality (\ref{f.4}) for $\overline{\sigma}_{n-1}(x)=-1$, $\widetilde{\sigma}_{n-1}(x)=0$ and $\widehat{\sigma}_{n-1}(x)=1$. Then we obtain

$$\left\{
\begin{array}{ll}
exp(h_{+1,x}-h_{0,x})=\prod_{y\in S(x)}\frac{\sum_{\omega_n(y)\in\Phi}exp\{J\beta\omega_n(y)+h_{\omega_n(y),y}\}}
{\sum_{\omega_n(y)\in\Phi}exp\{h_{\omega_n(y),y}\}},\\[2mm]
exp(h_{-1,x}-h_{0,x})=\prod_{y\in S(x)}\frac{\sum_{\omega_n(y)\in\Phi}exp\{-J\beta\omega_n(y)+h_{\omega_n(y),y}\}}
{\sum_{\omega_n(y)\in\Phi}exp\{h_{\omega_n(y),y}\}}.\\
\end{array}
\right.$$
Consequently
$$\left\{
\begin{array}{ll}
exp(h_{+1,x}-h_{0,x})=\prod_{y\in
S(x)}\frac{exp\{J\beta\}exp\{h_{+1,y}-h_{0,y}\}+exp\{-J\beta\}exp\{h_{-1,y}-h_{0,y}\}+1}
{exp\{h_{+1,y}-h_{0,y}\}+exp\{h_{-1,y}-h_{0,y}\}+1},\\[2mm]
exp(h_{-1,x}-h_{0,x})=\prod_{y\in
S(x)}\frac{exp\{-J\beta\}exp\{h_{+1,y}-h_{0,y}\}+exp\{J\beta\}exp\{h_{-1,y}-h_{0,y}\}+1}
{exp\{h_{+1,y}-h_{0,y}\}+exp\{h_{-1,y}-h_{0,y}\}+1}.\\
\end{array}
\right.$$
Hence, we can get (\ref{f.3}).

\textbf{Sufficiency.} Suppose that (\ref{f.3}) holds. It is equivalent to the representations
\begin{equation}\label{f.5}\prod_{y\in S(x)}\sum_{u\in\{-1,0,+1\}}exp(J\beta t u+h_{u, y})=
a(x)exp(h_{t, x}), \ \ t=-1, 0, +1 .
\end{equation}
for some function $a(x)>0, \ \ x\in V$. We have
\begin{equation}\label{f.6}
\textit{LHS of \ } (\ref{f.2})=\frac{1}{Z_n}exp(-\beta H(\sigma_{n-1})\prod_{x\in
W_{n-1}}\prod_{y\in S(x)}\sum_{u\in\{-1,0,+1\}}exp(J\beta
\sigma_{n-1}(x)u+h_{u, y}).
\end{equation}

Taking (\ref{f.5}) into account and denoting
$$A_n(x)=\prod_{x\in W_{n-1}}a(x),$$
from (\ref{f.6}) we get

\begin{equation}\label{f.7}
\textit{LHS of \ } (\ref{f.5})=\frac{A_{n-1}}{Z_n}exp(-\beta H(\sigma_{n-1}))\prod_{x\in W_{n-1}}exp(h_{\sigma_{n-1}(x),x}).
\end{equation}

Since $\mu^{(n)}, n\geq 1$ is probabilistic measure, then the following equation is true
$$\sum_{\sigma_{n-1}\in \Omega_{V_{n-1}}}\sum_{\omega_n\in \Omega_{W_n}}\mu^{(n)}(\sigma_{n-1}, \omega_n)=1.$$

Consequently from (\ref{f.7}) we obtain $Z_{n-1}A_{n-1}=Z_n$ and the validity of
(\ref{f.2}). The theorem is proved.\

\section{Translation-invariant Gibbs measures}\

Translation-invariant Gibbs measures corresponds to solutions
(\ref{f.3}) with  $z_{i,x}=z_i>0$ for all $x\in V$ and $i=-1,+1$. We introduce the notation
$z_{+1}=z_1, z_{-1}=z_2$ . Then (\ref{f.3}) has the form
\begin{equation}\label{f.8}\left\{
\begin{array}{ll}
z_1=\left(\frac{\lambda z_1+{1\over \lambda}z_2+1}{z_1+z_2+1}\right)^k, \\[2mm]
z_2=\left(\frac{{1\over \lambda}z_1+\lambda z_2+1}{z_1+z_2+1}\right)^k.
\end{array}
\right.
\end{equation}
We subtract the second equation in system (\ref{f.8}) from the
first, and we shall have
\begin{equation}\label{f.10}
(z_1-z_2)\left[1-\frac{(\lambda-\frac{1}{\lambda})((\lambda z_1+\frac{1}{\lambda}z_2+1)^{k-1}+\cdots+
(\frac{1}{\lambda}z_1+\lambda z_2+1)^{k-1})}{(z_1+z_2+1)^k}\right]=0.
\end{equation}

Hence $z_1=z_2$ or
$$(z_1+z_2+1)^k=\left(\lambda-\frac{1}{\lambda}\right)\left[\left(\lambda
z_1+\frac{1}{\lambda}z_2+1\right)^{k-1}+\cdots+\left(\frac{1}{\lambda}z_1+\lambda
z_2+1\right)^{k-1}\right]. $$

We consider the case $z_1=z_2=z$. In this case from (\ref{f.8}) we obtain
\begin{equation}\label{f.9}
z=\left(\frac{(\lambda+\frac{1}{\lambda})z+1}{2z+1}\right)^k.
\end{equation}

For solutions of the last equation the next proposition is hold.

\textbf{Proposition 1.} If $\overline{z}$ is the solution of the equation (\ref{f.9}), then
$$1\leq \overline{z}<\left({\lambda+{1\over\lambda}\over 2}\right)^k,$$
and $\overline{z}=1$ for $\lambda=1$.

The proof of Proposition 1 is obtained directly from the equation  (\ref{f.9}).

\textbf{Proposition 2.} \emph{For}  $k\geq2$ \emph{and for any values} $\lambda>0$
\emph{the equation (\ref{f.9}) has a unique positive solution.}

\textbf{Proof.} The proof will be carried out in three steps.

\textbf{Step 1.} Denoting $\sqrt[k]{z}=x$ the equation (\ref{f.9}) we rewrite in form
\begin{equation}\label{f.91}
\varphi(x)=2x^{k+1}-ax^k+x-1=0,
\end{equation}
where $a=\lambda+{{1}\over {\lambda}}\geq2$. Then the inequality from Proposition 1 has the form $1\leq x<{{a}\over{2}}$.

If $a=2$ (Є.х. $\lambda=1$) then the equation (\ref{f.91}) (the equation (\ref{f.9})) has a unique solution $x=1$ ($z=1$). Therefore we consider the case $a>2$ ($\lambda\neq1$).

By Proposition 1 it is clear that $1\leq x<{{a}\over{2}}$. Note that $\varphi(1)=2-a<0$ and $\varphi({a\over2})=1>0$, i.e. the equation (\ref{f.91}) has at least one positive solution for $1\leq x<{{a}\over{2}}$. Moreover, since the number of sign changes of the polynomial $\varphi(x)=2x^{k+1}-ax^k+x-1$ is three it follows from the known Descartes theorem on the number of positive roots of a polynomial (\cite{PV}, Corollary 1, pp. 39) that the equation (\ref{f.91}) has at most three positive solutions.

\textbf{Step 2.} In the second step of proof we use the Jacobi method for estimating the number of roots of a polynomial between $\alpha$ and $\beta$ (\cite{PV}, Remark, pp. 39). To do this, we make a substitution
$$y={{x-1}\over {{a\over 2}-x}} \ \left(\mbox {i.e.} \ x={1+{{a\over 2}}y\over1+y}\right)$$
and consider the polynomial
$$(1+y)^{k+1}\varphi\left({1+{{a\over 2}}y\over1+y}\right)=(a-2)\left[{y\over2}(y+1)^k-\left({a\over2}y+1\right)^k\right]=$$
$$=(a-2)\left[{1\over2}y^{k+1}+\left({1\over2}C_k^1-{a^k\over 2^k}\right)y^k+\left({1\over2}C_k^2-C_k^1{a^{k-1}\over 2^{k-1}}\right)y^{k-1}+\ldots+\left({1\over2}-C_k^{k-1}{a\over 2}\right)y-1\right]=$$
$$=(a-2)\left({1\over2}y^{k+1}+b_0y^k+b_1y^{k-1}+\ldots+b_{k-1}y+b_k\right)=(a-2)\psi(y).$$
Here
$$b_i={1\over2}C_k^{i+1}-C_k^i\left({a\over 2}\right)^{k-i}, \ i=0,1,2, \ldots, k-1, \ b_k=-1.$$
By the Jacobi method the number of positive roots of the polynomial $\psi(y)$ is the number of positive roots of the polynomial $\varphi(x)$ for $[1, {a\over2})$.

We note that if $b_i<0$ for all $i=1, 2, \ldots, k-1$ ($b_k=-1<0$) then independently on the sign of $b_0$ by the Descartes theorem the polynomial $\psi(y)$ has a unique positive solution. Thus we consider the case $i\neq0$.

If $b_i>0$, then
$$a<2\sqrt[k-i]{{k-i}\over2(i+1)}=t_1$$
ш $i<{k-2\over3}=i_0$, $i\in \{1, 2, \ldots, k-1\}$. Indeed, solving the inequality $b_i>0$ for $a$ the inequality $a<t_1$ is obtained directly. On the other hand, the inequality $b_i>0$ is equivalent to the inequality
$${1\over2}C_k^{i+1}>C_k^i\left({a\over 2}\right)^{k-i}.$$
From this inequality we get 
$${k-i\over2(i+1)}>\left({a\over 2}\right)^{k-i},$$
here the right side is greater than one. Hence we have
$$i<{k-2\over3}=i_0.$$
Consequently $b_i<0$ for any $i\geq i_0.$

\textbf{Step 3.} In this step we prove that if $b_i>0$ for $0\neq i<i_0$ then $b_{i-1}$ is also positive. We suppose $b_i>0$ but $b_{i-1}<0$. If $b_i>0$ then it is already known
$$a<2\sqrt[k-i]{{k-i}\over2(i+1)}=t_1.$$
From $b_{i-1}<0$ we have
$$a>2\sqrt[k-i+1]{{k-i+1\over 2i}}=t_2.$$

We prove that $t_1<t_2$. Indeed, $t_1<t_2$ is equivalent to the inequality
$$\left({k-i\over2(i+1)}\right)^{k-i+1}<\left({k-i+1\over2i}\right)^{k-i}.$$
Denoting $k-i=n, 1\leq n< k$ (since $i\neq0$ here $n\neq k$), we rewrite the last inequality
\begin{equation}\label{f.92}
\left({n\over2(k-n+1)}\right)^{n+1}<\left({n+1\over2(k-n)}\right)^n.
\end{equation}

Using mathematical induction we prove the inequality (\ref{f.92}). For $n=1$ we obtain the inequality $4k^2-k+1>0$ which is true for any values $k$. We suppose that (\ref{f.92}) is hold for $n$. We prove the inequality
$$\left({n+1\over2(k-n)}\right)^{n+2}<\left({n+2\over2(k-n-1)}\right)^{n+1}.$$
We transform and estimate the left-hand side of the last inequality
$$\left({n+1\over2(k-n)}\right)^{n+2}=\left({n+1\over2(k-n)}\right)^{n+2}\cdot\left({n\over2(k-n+1)}\right)^{n+1}\cdot\left({2(k-n+1)\over n}\right)^{n+1}=$$
$$=\left({n\over2(k-n+1)}\right)^{n+1}\cdot\left({n+1\over2(k-n)}\right)^{n+2}\cdot\left({2(k-n+1)\over n}\right)^{n+1}<$$
$$<\left({n+1\over2(k-n)}\right)^n\cdot\left({n+1\over2(k-n)}\right)^{n+2}\cdot\left({2(k-n+1)\over n}\right)^{n+1}=$$
$$=\left({n+1\over2(k-n)}\right)^n\cdot\left({n+1\over2(k-n)}\right)^{n+2}\cdot\left({2(k-n+1)\over n}\right)^{n+1}\cdot\left({n+2\over2(k-n-1)}\right)^{n+1}\cdot\left({2(k-n-1)\over n+2}\right)^{n+1}=$$
$$=\left({n+2\over2(k-n-1)}\right)^{n+1}\cdot\left({n+1\over2(k-n)}\right)^{2n+2}\cdot\left({2(k-n+1)2(k-n-1)\over n(n+2)}\right)^{n+1}.$$
Consequently it is necessary to prove the inequality
$$\left({n+2\over2(k-n-1)}\right)^{n+1}\cdot\left({n+1\over2(k-n)}\right)^{2n+2}\cdot\left({2(k-n+1)2(k-n-1)\over n(n+2)}\right)^{n+1}<\left({n+2\over2(k-n-1)}\right)^{n+1},$$
which is equivalent to the inequality
$$\left({n+1\over2(k-n)}\right)^{2n+2}<\left({n(n+2)\over 4((k-n)^2-1)}\right)^{n+1}.$$
From the last inequality we obtain $i<{k+1\over2}$. Since $i<{k-2\over3}$ and ${k-2\over3}<{k+1\over2}$ the inequality $i<{k+1\over2}$ is hold. Hence the equation (\ref{f.9}) has a unique solution for any values $\lambda>0$ and $k\geq2$. The proposition is proved.\

In the case $z_1=z_2=z$ by Proposition 2 we get that the system (\ref{f.8}) has
a unique solution $(z^*,z^*)$ for $\lambda>0$ and $k\geq2$.

The following theorem holds.

\textbf{Theorem 2.}  \emph{Let $k=2.$ Then for the Blum-Kapel model there exist
$\lambda_{cr}\approx {2.1132163}$ such that there exist
one translation-invariant Gibbs measure $\mu_0$ for $0<\lambda\leq\lambda_{cr}$ and there are exactly three translation-invariant Gibbs measures $\mu_0, \mu_1, \mu_2$ for  $\lambda>\lambda_{cr}$.}

\textbf{Proof.} In the case $k=2$ from (\ref{f.10}) we get
$$(z_1-z_2)\cdot\left[1-\frac{(\lambda-\frac{1}{\lambda})((\lambda+\frac{1}{\lambda})(z_1+z_2)+2)}{(z_1+z_2+1)^2}\right]=0.$$

In the case $z_1=z_2$ it is already known that there is a unique solution for any $\lambda>0$.

Let $z_1\neq z_2$. Then
$$(z_1+z_2+1)^2=\left(\lambda-\frac{1}{\lambda}\right)\cdot\left[\left(\lambda+\frac{1}{\lambda}\right)(z_1+z_2)+2\right].$$
This equation is equivalent to the equation for $(z_1+z_2)$
$$(z_1+z_2)^2-\left(\lambda^2-\frac{1}{\lambda^2}-2\right)(z_1+z_2)+1-2\left(\lambda-\frac{1}{\lambda}\right)=0,$$
which solutions has form
$$(z_1+z_2)_{1,2}=\frac{\lambda^4-2\lambda^2-1\pm\sqrt{D}}{2\lambda^2}=\varphi_{1,2}(\lambda),$$
where
$$D=(\lambda+1)(\lambda-1)^2(\lambda^5+\lambda^4-2\lambda^3+6\lambda^2+\lambda+1)\geq0$$
for any $\lambda>0$.

It is not difficult to show that
$$\varphi_1(\lambda)=\frac{\lambda^4-2\lambda^2-1-\sqrt{D}}{2\lambda^2}<0$$
for any $\lambda>0$ and
$$\varphi_2(\lambda)=\frac{\lambda^4-2\lambda^2-1+\sqrt{D}}{2\lambda^2}>0$$
for $\lambda>{1+\sqrt{17}\over4}\approx1.28078.$

Thus $z_1+z_2=\varphi_2(\lambda).$ From the system of equations (\ref{f.8}) we obtain
$$(z_1+z_2)(z_1+z_2+1)^2=\left(\lambda^2+\frac{1}{\lambda^2}\right)(z_1+z_2)^2+2\left(\lambda+\frac{1}{\lambda}\right)
(z_1+z_2)+2\left(2-\left(\lambda^2+\frac{1}{\lambda^2}\right)\right)z_1z_2+2.$$
In respect that $z_1+z_2=\varphi_2(\lambda)$ we have the quadratic equation for $z_1$:

$$2\left(2-\left(\lambda^2+\frac{1}{\lambda^2}\right)\right)z_1^2-2\left(2-\left(\lambda^2+\frac{1}{\lambda^2}\right)\right)\varphi_2(\lambda) z_1-$$
\begin{equation}\label{f.11}
-\left[\left(\lambda^2+\frac{1}{\lambda^2}\right)\varphi_2^2(\lambda)+2\left(\lambda+\frac{1}{\lambda}\right)\varphi_2(\lambda)-
\varphi_2(\lambda)(\varphi_2(\lambda)+1)^2+2\right]=0,
\end{equation}
Discriminant of this quadratic equation is
$$D_1=2^2\left(2-\left(\lambda^2+\frac{1}{\lambda^2}\right)\right)^2\varphi_2^2(\lambda)+8\left(2-\left(\lambda^2+\frac{1}
{\lambda^2}\right)\right)\times$$
$$\times\left[\left(\lambda^2+\frac{1}{\lambda^2}\right)\varphi_2^2(\lambda)+2\left(\lambda+\frac{1}{\lambda}\right)\varphi_2(\lambda)-
\varphi_2(\lambda)(\varphi_2(\lambda)+1)^2+2\right]>0$$
for $\lambda>\lambda_{cr}\approx2.1132163.$  Then the equation (\ref{f.11}) has two positive solutions for $\lambda>\lambda_{cr}$:

$$z_1^{(1)}(\lambda)={1\over
2}\varphi_2(\lambda)+\frac{\sqrt{D_1}}{4(\lambda-{1\over\lambda})^2},
\\ \  \  \ z_1^{(2)}(\lambda)={1\over
2}\varphi_2(\lambda)-\frac{\sqrt{D_1}}{4(\lambda-{1\over\lambda})^2}.
$$

Computer and cumbersome analysis shows that
$$\lim_{\lambda\rightarrow+\infty}z_1^{(1)}(\lambda)=+\infty, \ \lim_{\lambda\rightarrow+\infty}z_1^{(2)}(\lambda)=0, \ \lim_{\lambda\rightarrow\lambda_{cr}}z_1^{(1)}(\lambda)=\lim_{\lambda\rightarrow\lambda_{cr}}z_1^{(2)}(\lambda)=
{1\over2}\varphi_2(\lambda_{cr})\approx1.487$$
and $z_1^{(1)}>0$, $z_1^{(2)}>0$ (see Fig.1).

In addition, from the notation $z_1+z_2=\varphi_2(\lambda)$ we have
$z_2^{(1)}=z_1^{(2)}$, $z_1^{(1)}=z_2^{(2)}$, i.e. solutions of the system of equations (\ref{f.8})
are symmetric: $(z_1,z_2)$ and $(z_2,z_1)$.

It is known from the Proposition 2 that the system of equations (\ref{f.8}) has a unique positive solution for $k\geq2, \ \lambda>0$ and $z_1=z_2=z^*$. In particular we can find explicit form of this solution for $k=2$. For this we consider the equation (\ref{f.9}) for $k=2$:
\begin{equation}\label{f.12}
z=\left(\frac{(\lambda+\frac{1}{\lambda})z+1}{2z+1}\right)^2,
\end{equation}
which is equivalent to the equation
$$g(z)=4z^3+(4-a^2)z^2+(1-2a)z-1=0.$$
Using the Cardano formula we find solution of the last equation:
\begin{equation}\label{f.13}
z^*=\frac{1}{12\lambda^2}\left(\sqrt[3]{A+6\lambda^4\sqrt{{3B\over \lambda}}}+{C\over \sqrt[3]{A+6\lambda^4\sqrt{{3B\over\lambda}}}}+(\lambda^2-1)^2\right),
\end{equation}
where
$$A=\lambda^{12}-6\lambda^{10}+36\lambda^9-3\lambda^8-36\lambda^7+232\lambda^6-36\lambda^5-3\lambda^4+36\lambda^3-6\lambda^2+1,$$
$$B=4\lambda^{10}-17\lambda^9+4\lambda^8+188\lambda^7-616\lambda^6+874\lambda^5-616\lambda^4+188\lambda^3+4\lambda^2-17\lambda+4,$$
$$C=\lambda^8-4\lambda^6+24\lambda^5-6\lambda^4+24\lambda^3-4\lambda^2+1.$$
Thus, for $0<\lambda\leq\lambda_{cr}$ there is a unique translation-invariant Gibbs measure $\mu_0$, corresponding to unique solution $(z^*, z^*)$ of the system of equations (\ref{f.8}) and  for $\lambda>\lambda_{cr}$ there are three translation-invariant Gibbs measures $\mu_0, \mu_1, \mu_2$, corresponding to solutions $(z^*, z^*)$, $(z_1,z_2)$ and $(z_2,z_1)$, respectively. The theorem is proved.\

\begin{center}
\includegraphics[width=8cm]{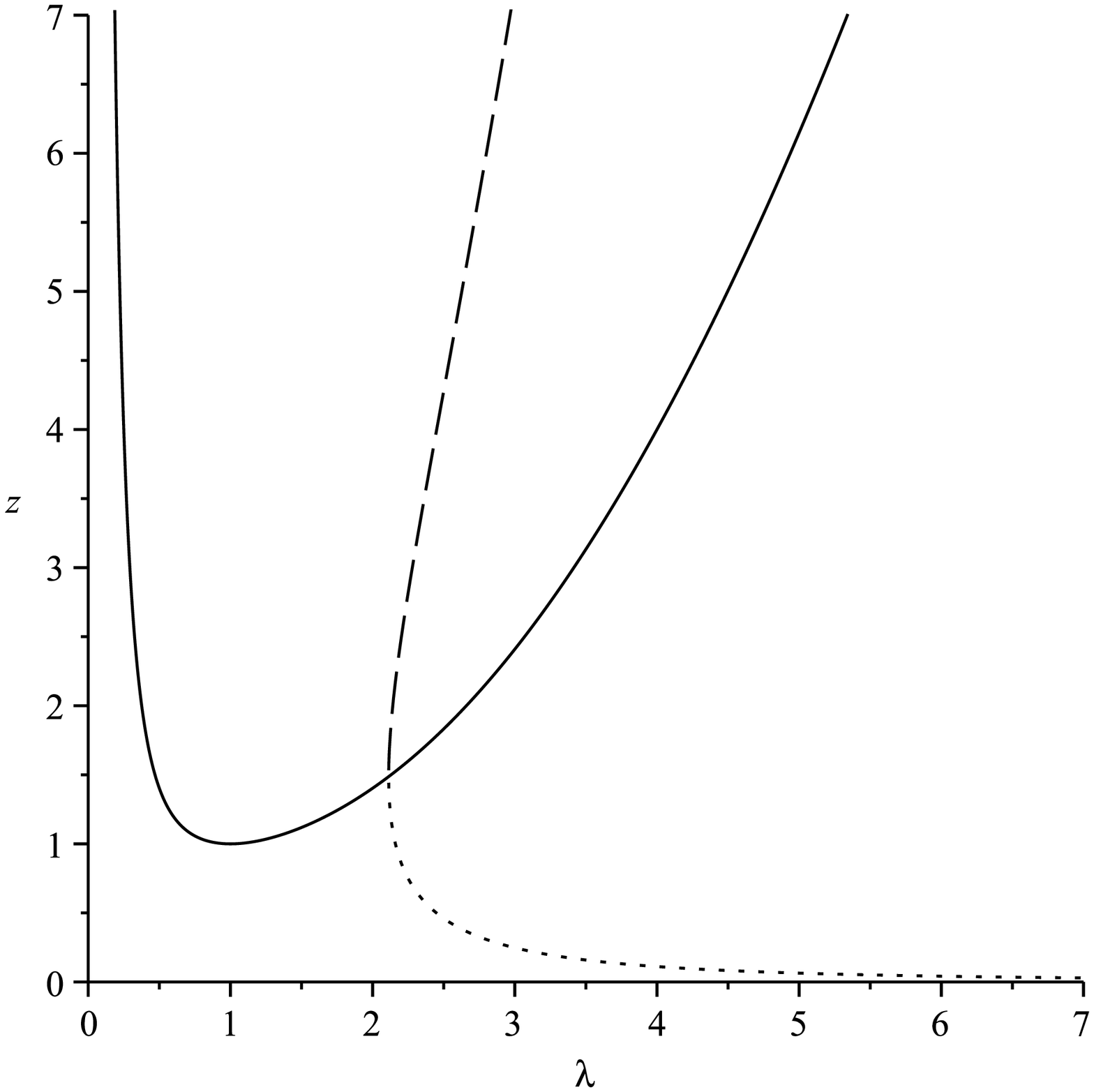}
\end{center}
\begin{center}{\footnotesize \noindent
 Fig.~1.
Graph of the functions $z^*(\lambda)$(continuous curve), $z_1(\lambda)$ (shaded curve), $z_2(\lambda)$ (pointwise curve)}. \\
\end{center}

\textbf{Remark 1.} Since $\lambda=\exp({J\over T})$, where $T>0$ is temperature then $T_{cr}={J\over \ln \lambda_{cr}}$ and by Theorem 2 for the Blum-Kapel model there is a unique translation-invariant Gibbs measure $\mu_0$ for $T\geq T_{cr}$, and there are exactly three translation-invariant Gibbs measures $\mu_0, \mu_1, \mu_2$ for $0<T< T_{cr}$.

\section{Extremality of the measure $\mu_0$}\

In this section we study the extremality of the measure $\mu_0$, corresponding to the solution $(z^*, z^*)$.
To check the extremality of the Gibbs measure, we apply arguments of a
reconstruction on trees from \cite{FK} and methods from \cite{Ke}, \cite{MSW}. We consider
Markov chain with states $\{-1,0,1\}$ and transition probabilities matrix $\mathbb P=(P_{ij})$.

$$\mu^{(n)}=\frac{1}{Z}\prod_{x\in W_n} exp\{-J\beta x\sigma (x)+h_{\sigma (x)}\},$$

$$P_{\sigma (x) \sigma (y)}=\frac{exp\{-J\beta \sigma (x) \sigma (y)+h_{\sigma (y)}\}}{\sum_{\sigma (y)\in\{-1, 0, +1\}}exp\{-J\beta \sigma (x) \sigma (y) +h_{\sigma (y)}\}}.$$

Hence, using $z'_{i, x}=\frac{z_{i, x}}{z_{0, x}}, i=1,2$,
we get
$$P_{-1, -1}=\frac{\lambda^2z'_1}{\lambda^2z'_1+\lambda+z'_2};   \ \ \ P_{-1, 0}=\frac{\lambda}{\lambda^2z'_1+\lambda+z'_2}; \ \ \  P_{-1, +1}=\frac{z'_2}{\lambda^2z'_1+\lambda+z'_2};$$
$$P_{0, -1}=\frac{z'_1}{z'_1+1+z'_2};   \ \ \ P_{0, 0}=\frac{1}{z'_1+1+z'_2}; \ \ \  P_{0, +1}=\frac{z'_2}{z'_1+1+z'_2};$$
$$P_{+1, -1}=\frac{z'_1}{z'_1+\lambda+\lambda^2z'_2};   \ \ \ P_{+1, 0}=\frac{\lambda}{z'_1+\lambda+\lambda^2z'_2}; \ \ \  P_{+1, +1}=\frac{\lambda^2z'_2}{z'_1+\lambda+\lambda^2z'_2}.$$
Consequently (we set $z'_i=z_i$ in what follows)

$$\mathbb{P}=\left(%
\begin{array}{ccc}
  \frac{\lambda^2z_1}{\lambda^2z_1+\lambda+z_2} & \frac{\lambda}{\lambda^2z_1+\lambda+z_2} & \frac{z_2}{\lambda^2z_1+\lambda+z_2} \\
  \frac{z_1}{z_1+1+z_2} & \frac{1}{z_1+1+z_2} & \frac{z_2}{z_1+1+z_2} \\
  \frac{z_1}{z_1+\lambda+\lambda^2z_2} & \frac{\lambda}{z_1+\lambda+\lambda^2z_2} & \frac{\lambda^2z_2}{z_1+\lambda+\lambda^2z_2} \\
\end{array}%
\right). $$

For considered solution $\mathbb{P}$  the matrix has the form
$(z_1=z_2=z)$ :

$$\mathbb{P}=\left(%
\begin{array}{ccc}
  \frac{\lambda^2z}{\lambda^2z+\lambda+z} & \frac{\lambda}{\lambda^2z+\lambda+z} & \frac{z}{\lambda^2z+\lambda+z} \\
  \frac{z}{2z+1} & \frac{1}{2z+1} & \frac{z}{2z+1} \\
  \frac{z}{z+\lambda+\lambda^2z} & \frac{\lambda}{z+\lambda+\lambda^2z} & \frac{\lambda^2z}{z+\lambda+\lambda^2z} \\
\end{array}%
\right). $$

\subsection{Conditions for non-extremality of the measure $\mu_0$}\

It is known that a sufficient condition (i.e., the
Kesten-Stigum condition) for non-extremality of a Gibbs measure $\mu$ corresponding to the matrix $\mathbb P$ is that $k\lambda_2^2>1$, where $\lambda_2$ is the second largest (in absolute value) eigenvalue of $\mathbb P$ (see \cite{Ke}).

We shall find conditions of non-extremality of the measure
corresponding to a unique solution $(z^*, z^*) (z^*=z)$.
It is clear that the eigenvalues of this matrix are
$$s_1=\frac{(\lambda-1)^2z}{((\lambda^2+1)z+\lambda)(2\lambda+1)}, \ \  \ s_2=\frac{(\lambda^2-1)z}{(\lambda^2+1)z+\lambda}, \ \ s_3=1,$$
where $z$ is the solution (\ref{f.12}). We find $\max\{\mid s_1\mid,  \mid
s_2\mid\}$:

$$\mid s_1\mid-\mid s_2\mid=\frac{(\lambda-1)^2z}{((\lambda^2+1)z+\lambda)(2\lambda+1)}-\frac{\mid\lambda-1\mid (\lambda+1)z}{(\lambda^2+1)z+\lambda}$$
Let $\lambda>1$ then

$$\mid s_1\mid-\mid s_2\mid=\frac{2(1-\lambda)(\lambda^2+\lambda+1)z}{((\lambda^2+1)z+\lambda)(2\lambda+1)}<0.$$
For $\lambda<1$
$$\mid s_1\mid-\mid s_2\mid=\frac{2\lambda(\lambda-1)(\lambda+2)z}{((\lambda^2+1)z+\lambda)(2\lambda+1)}<0.$$
Then for any $\lambda>0$ we have
$$\max\{\mid s_1\mid, \mid s_2\mid\}=\mid s_2\mid.$$

Consequently $s_1<\mid s_2\mid<s_3=1$.

Now we check the Kesten-Stigum condition  for non-extremality of the measure $\mu_0$: $2 {s_2}^2>1$, i.e.
$$2{s_2}^2-1=2\cdot\left(\frac{(\lambda^2-1)z}{(\lambda^2+1)z+\lambda}\right)^2-1>0,$$
where $z$ has the form (\ref{f.13}). Using Maple one can see that
the last inequality holds for $\lambda\in (0,
\lambda_1)\cup(\lambda_2, +\infty)$, where $\lambda_1\approx0.336135$ ш $\lambda_2\approx2.975$, i.e. the measure $\mu_0$ is non-extreme under this condition (see Fig.2).

\begin{center}
\includegraphics[width=8cm]{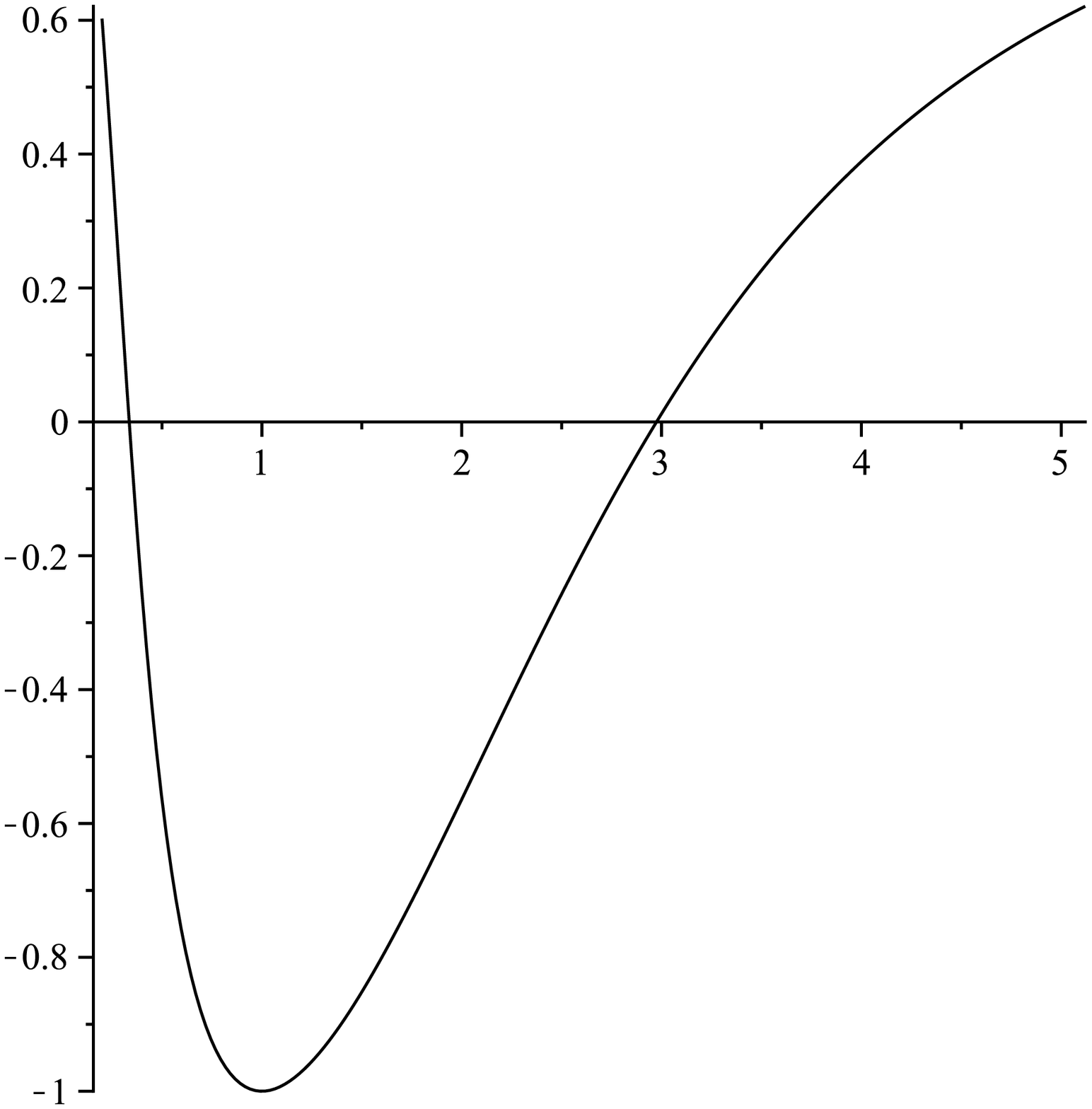}
\end{center}
\begin{center}{\footnotesize \noindent
 Fig.~2.
Graph of the function $2s_2^2-1$}. \\
\end{center}

Thus the following theorem holds.

\textbf{Theorem 3.} Let $k=2, \lambda\in (0, \lambda_1)\cup(\lambda_2,
+\infty)$, where $\lambda_1\approx0.336135$ and $\lambda_2\approx2.975$. Then for the Blum-Kapel model the measure $\mu_0$ is non-extreme.\

\textbf{Remark 2.} We note that $T={J\over \ln \lambda}$, where $T>0$ is temperature and since $T_1={J\over \ln \lambda_{1}}<0$ then in the case $k=2$ the measure $\mu_0$ is non-extreme for $T\in (0, T_2)$.

\subsection{Conditions for extremality of the measure $\mu_0$.}

If we from a Cayley tree $\Gamma^k$ remove an arbitrary edge $\langle x^0, x^1\rangle=l\in L$, then it is divided into two components $\Gamma^k_{x^0}$ and $\Gamma^k_{x^1}$ each of which is called semi-infinite Cayley tree or Cayley subtree.

Let us first give some necessary definitions from \cite{MSW}. We consider finite complete subtrees $\mathcal T$ that are initial points of Cayley tree $\Gamma^k_{x^0}$. The boundary $\partial \mathcal T$ of subtree $\mathcal T$ consists the neighbors which are on $\Gamma^k_{x^0}\setminus \mathcal T$. We
identify subgraphs of $\mathcal T$ with their vertex sets and write $E(A)$ for the edges within a subset $A$ and $\partial A$.

In \cite{MSW} the key ingredients are two quantities $\kappa$ and $\gamma$. Both are properties of the
collection of Gibbs measures $\{\mu^\tau_{{\mathcal T}}\}$, where the boundary condition $\tau$ is fixed and $\mathcal T$ ranges
over all initial finite complete subtrees of $\Gamma^k_{x^0}$. For a given subtree $\mathcal T$ of $\Gamma^k_{x^0}$ and a vertex $x\in\mathcal T$ we write $\mathcal T_x$ for the (maximal) subtree of $\mathcal T$  rooted at $x$. When $x$ is not the root of $\mathcal T$, let $\mu_{\mathcal T_x}^s$ denote the (finite-volume) Gibbs measure in which the parent of $x$ has its spin fixed to $s$ and the configuration on the bottom boundary of ${\mathcal T}_x$ (i.e., on $\partial {\mathcal T}_x\setminus \{\mbox{parent of}\ \ x\}$) is specified by $\tau$.

For two measures $\mu_1$ and $\mu_2$ on $\Omega$, $\|\mu_1-\mu_2\|_x$ denotes the variation distance between the projections of $\mu_1$ and $\mu_2$ onto the spin at $x$, i.e.,
$$\parallel\mu_1-\mu_2\parallel_x=\frac{1}{2}\sum_{i\in\{-1, 0, +1\}}\mid\mu_1(\sigma(x)=i)-\mu_2(\sigma(x)=i)\mid.$$

Let $\eta^{x, s}$ be the configuration $\eta$ with the spin at $x$ set to $s.$

Following (\cite{MSW}) define
$$\kappa\equiv\kappa(\mu)=\sup_{x\in\Gamma^k}\max_{x, s, s'}\parallel\mu_{\mathcal T_x}^s-\mu_{\mathcal T_x}^{s'}\parallel_x,$$
$$\gamma\equiv\gamma(\mu)=\sup_{A\subset\Gamma^k}\max\parallel\mu_A^{\eta^{y, s}}-\mu_A^{\eta^{y, s'}}\parallel_x,$$
where the maximum is taken over all boundary conditions  $\eta$, all sites
$y\in\partial A$, all neighbors $x\in A$  of $y$ and all spins $s, s'\in\{-1, 0, +1\}.$\

It is known that a sufficient condition for extremality of the translation-invariant Gibbs measure is $k\kappa\gamma<1$ (see \cite{MSW}, Theorem 9.3).

Note that $\kappa$ has the particularly simple form
$$k=\frac{1}{2}\max{\sum_{l\in\{-1, 0, +1\}}\mid P_{il}-P_{jl}\mid}.$$

Hence, it is clearly that $\mid P_{il}-P_{jl}\mid=0$ for $i=j.$ Using methods from \cite{MSW} we compute (for $i\neq j$)
$$\sum_{l\in\{-1, 0, +1\}}\mid P_{il}-P_{jl}\mid=\left\{%
\begin{array}{ll}
    \frac{((\lambda+1)(2z+1)+\mid \lambda-1\mid)\mid\lambda-1\mid z}{(\lambda^2z+z+\lambda)(2z+1)}, & \hbox{$i=-1, j=0$ or $i=0, j=-1,$}
    \\
    \frac{2\mid\lambda^2-1\mid z}{\lambda^2z+z+\lambda}, & \hbox{$i=-1, j=+1$ or $i=+1, j=-1,$}
   \\

    \frac{((\lambda+1)(2z+1)+\mid \lambda-1\mid)\mid\lambda-1\mid z}{(\lambda^2z+z+\lambda)(2z+1)}, & \hbox{$i=0, j=+1$ or $i=+1, j=0.$} \\
\end{array}%
\right.    $$

We note that
$$\kappa=\frac{\mid\lambda^2-1\mid z}{\lambda^2z+z+\lambda}.$$

Now, similarly to the work (\cite{MSW}, p.15) we shall find the estimate for $\gamma$
in the following form:
$$\gamma=\max\{\parallel\mu_A^{\eta^{y,-1}}-\mu_A^{\eta^{y,
0}}\parallel_x, \parallel\mu_A^{\eta^{y,-1}}-\mu_A^{\eta^{y,
+1}}\parallel_x, \parallel\mu_A^{\eta^{y,0}}-\mu_A^{\eta^{y,
+1}}\parallel_x\},$$
where

$$\parallel\mu_A^{\eta^{y,-1}}-\mu_A^{\eta^{y,
0}}\parallel_x=\frac{1}{2}\sum_{s\in\{-1, 0,
+1\}}\mid\mu_A^{\eta^{y, -1}}(\sigma(x)=s)-\mu_A^{\eta^{y,
0}}(\sigma(x)=s)\mid=$$

$$=\frac{1}{2}(\mid P_{-1, -1}-P_{0,
-1}\mid +\mid P_{-1, 0}-P_{0, 0}\mid +\mid P_{-1, +1}-P_{0,
+1}\mid)=$$

$$=\frac{1}{2}\frac{((\lambda+1)(2z+1)+\mid\lambda-1\mid)\mid\lambda-1\mid
z}{(\lambda^2z+z+\lambda)(2z+1)}\leq\frac{\mid\lambda^2-1\mid
z}{\lambda^2z+z+\lambda},$$

$$\parallel\mu_A^{\eta^{y,-1}}-\mu_A^{\eta^{y,
+1}}\parallel_x=\frac{1}{2}\sum_{l\in\{-1, 0, +1\}}\mid P_{-1,
l}-P_{+1, l}\mid=\frac{\mid\lambda^2-1\mid
z}{\lambda^2z+z+\lambda},$$

$$\parallel\mu_A^{\eta^{y,0}}-\mu_A^{\eta^{y,
+1}}\parallel_x=\frac{1}{2}\sum_{l\in\{-1, 0, +1\}}\mid P_{0,
l}-P_{+1, l}\mid=$$

$$=\frac{1}{2}\frac{((\lambda+1)(2z+1)+\mid\lambda-1\mid)\mid\lambda-1\mid
z}{(\lambda^2z+z+\lambda)(2z+1)}\leq\frac{\mid\lambda^2-1\mid
z}{\lambda^2z+z+\lambda}.$$

Consequently
$$\gamma\leq\frac{\mid\lambda^2-1\mid z}{\lambda^2z+z+\lambda}.$$

We check the condition $2\kappa\gamma<1$ for $\mu_0$ which is equivalent to the inequality
$$(\lambda^4-6\lambda^2+1)z^2-2\lambda(\lambda^2+1)z-\lambda^2<0$$
where $z$ is defined by (\ref{f.13}). Using computer analysis we obtain that the last inequality holds for $\lambda_1<\lambda<\lambda_2$, where $\lambda_1\approx0.336135$ ш $\lambda_2\approx2.975$ (see Fig.3).

\begin{center}
 \includegraphics[width=8cm]{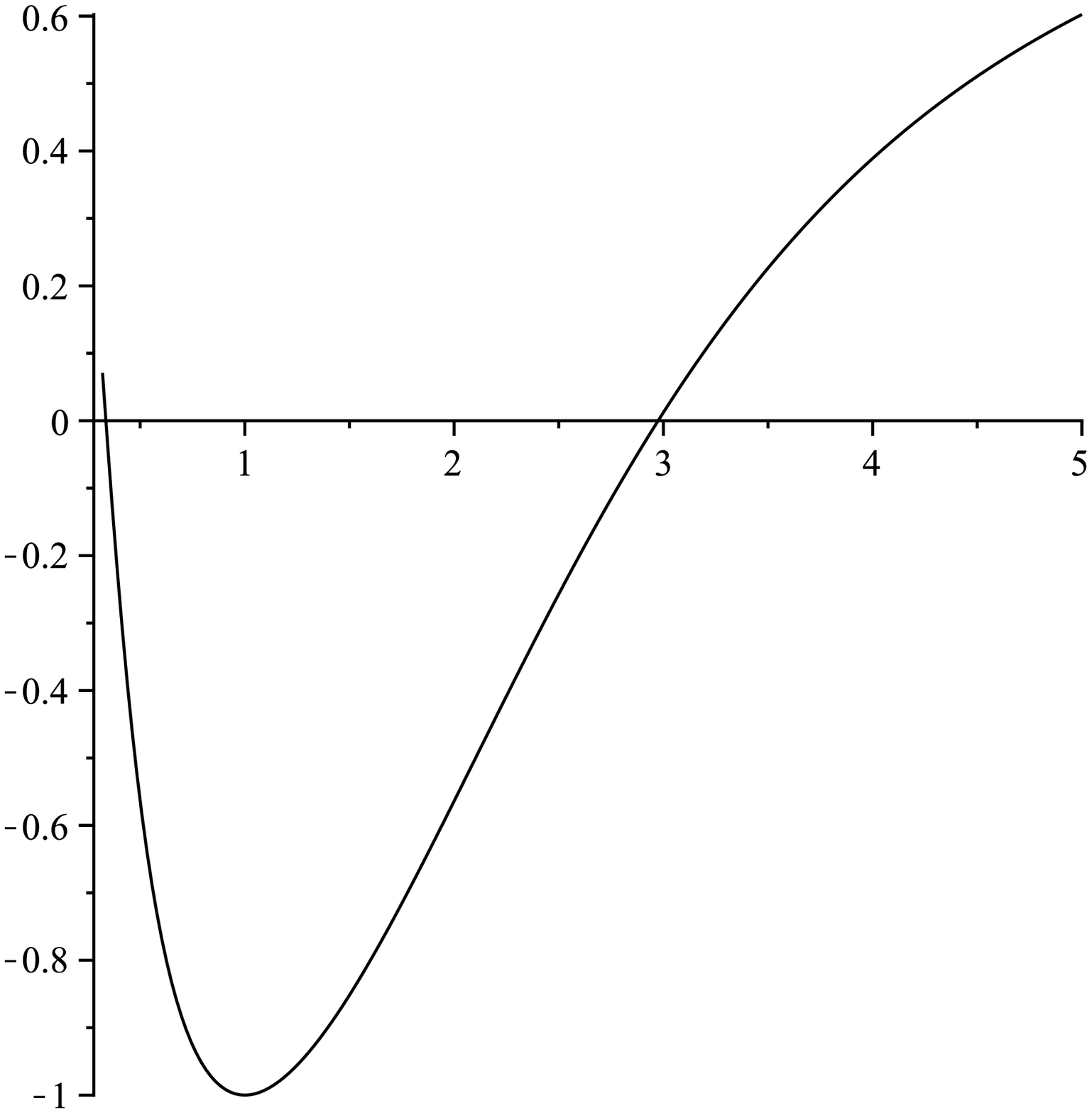}
\end{center}
\begin{center}{\footnotesize \noindent
 Fig.~3.
 Graph of the function $2\kappa\gamma-1.$}\
\end{center}

Thus the following theorem is true.

\textbf{Theorem 4.} Let $k=2$. Then for the Blum-Kapel model the measure
$\mu_0$ is extreme for $\lambda_1<\lambda<\lambda_2$.

\textbf{Remark 3.} Since $T_1<0$ then it follows from Remark 2 and Theorem 4 that in the case $k=2$ the measure $\mu_0$ is extreme for $T>T_2$.

\textbf{Remark 4.} To check (not) extremality of measures $\mu_1, \mu_2$ is very difficult even  with the help of computer analysis. Therefore this problem remains open.\

Since the set of all limit Gibbs measures forms a nonempty convex compact subset of the set of all probability
measures (\cite{6}-\cite{Si}) then the following theorem is true.

\textbf{Theorem 5.} If $k=2$ and $\lambda_{cr}<\lambda<\lambda_2$ (i.e. for $0<T<T_{cr}$ ш $T>T_2$) then there are at least two extreme Gibbs
measures for the Blum-Kapel model.

\textbf{Proof.} By Theorem 2 it is known that if $0<\lambda\leq\lambda_{cr}$
then there is unique translation-invariant Gibbs measure $\mu_0$. By Theorem 4 if $\lambda_1<\lambda<\lambda_2$, then the measure $\mu_0$ is extreme. For $\lambda>\lambda_{cr}$ we have measure $\mu_0$ and at least two new
measures $\mu_1, \mu_2$ mentioned in Theorem 2. If we assume that all the new measures are not extreme in $(\lambda_{cr}, \lambda_2)$
then there remains only one known extreme measure $\mu_0$. But in this case the non-extreme
measures can not be decomposed only into the unique measure $\mu_0$. Consequently, for $\lambda_{cr}<\lambda<\lambda_2$ at least
one of the new measures must be extreme. The theorem is proved.\\

\textbf{Acknowledgments.} The authors are very grateful to Professor U. A. Rozikov for his useful advice.

\end{document}